# Hall coefficient and $H_{c2}$ in underdoped $LaFeAsO_{0.95}F_{0.05}$

Y. Kohama[1,2(a)], Y. Kamihara[3,4], S. Riggs[1], F. F. Balakirev[1], T. Atake[2], M. Jaime[1], M. Hirano[3,4] and H. Hosono[3,4]

[1]*MPA-NHMFL, Los Alamos National Laboratory, Los Alamos, New Mexico 87545*
[2]*Materials and Structures Laboratory, Tokyo Institute of Technology, Mail-Box: R3-7, 4259 Nagatsuta-cho, Midori-ku, Yokohama 226-8503, Japan*
[3]*ERATO-SORST, Japan Science and Technology Agency in Frontier Research Center, Tokyo Institute of Technology, Mail-Box: S2-13, 4259 Nagatsuta-cho, Midori-ku, Yokohama 226-8503, Japan*
[4]*Frontier Research Center, Tokyo Institute of Technology*
*Mail-Box: S2-13, 4259 Nagatsuta-cho, Midori-ku, Yokohama 226-8503, Japan*

Abstract

The electrical resistivity and Hall coefficient of $LaFeAsO_{0.95}F_{0.05}$ polycrystalline samples were measured in pulsed magnetic fields up to $\mu_0 H$ = 60 T from room temperature to 1.5 K. The resistance of the normal state shows a negative temperature coefficient ($d\rho/dT < 0$) below 70 K for this composition, indicating insulating ground state in underdoped LaFeAsO system in contrast to heavily doped compound. The charge carrier density obtained from Hall effect can be described as constant plus a thermally activated term with an energy gap $\Delta E$ = 630 K. Upper critical field, $H_{c2}$, estimated from resistivity measurements, exceeds 75 T with zero-field $T_c$ = 26.3 K, suggesting an unconventional nature for superconductivity.

## Introduction

The discovery of superconductivity in lanthanum iron oxyarsenide, $LaFeAsO_{1-x}F_x$ ($T_c$ ~ 30 K) [1,2], and its rare-earth analogs ($T_c$ = 40 ~ 50 K) [3-5] has prompted intense research aimed at determining the crystal structure, electronic properties, and the mechanism of superconductivity in this new superconductor. Although the properties of this family differ strongly from those of conventional Bardeen-Cooper-Schrieffer (BCS) superconductors, some similarities to the high-$T_c$ cuprates have been observed, such as the layered structure [1], extremely high upper critical field ($H_{c2}$) [6], and possibly a pseudogap [7]. Resistivity ($\rho$) and Hall coefficient ($R_H$) measurements in high magnetic field are sensitive to the electronic structure and a presence of pseudogap [8], and allow for the estimation of $H_{c2}$.

Much of the experimental efforts so far were focused on understanding the property above 10 % F-doped iron oxyarsenides [6,9,10], in which $R_H$ and $H_{c2}$ were reported. However the resistivity [1] and magnet susceptibility [1] in lightly doped sample are quite different from those of heavily doped samples, and dramatic changes in $R_H$ and $H_{c2}$ are expected. Hence, the transport measurements reported here are crucial for understanding the electronic properties as a function of band filling.

In this letter, we report results of resistivity and Hall coefficient measurements in polycrystalline samples of 5% F-doped iron oxyarsenide, $LaFeAsO_{0.95}F_{0.05}$, using a 60 T capacitor-driven pulsed magnets at the National High Magnetic Field Laboratory. We find a pronounced negative temperature coefficient ($d\rho/dT < 0$) in the low temperature resistance, a thermal activation behavior in the charge carrier density, and a $T^2$ dependence of Hall angle. $H_{c2}$ is estimated to be 75.3 T in $LaFeAsO_{0.95}F_{0.05}$, which is ~16 % greater than published results in 11% F-doped iron oxyarsenide, $LaFeAsO_{0.89}F_{0.11}$.

## Experiment

Polycrystalline $LaFeAsO_{0.95}F_{0.05}$ was prepared by solid-state reactions, as described elsewhere [1], and $x$ values were determined from Vegard's law [11]. The magnetic field dependence of electrical resistivity, Hall coefficient and bulk penetration were measured at fixed temperatures using a 60 T capacitor bank-driven pulsed magnet, with 100 msec pulse duration. Electrical resistivity and Hall coefficient were measured at 80 kHz, using a standard 6-contact Hall bar configuration. Bulk penetration measurements were carried out using a Tunnel Diode Oscillator (TDO) technique at 31 MHz [12].

## Results and Discussion

The result of our zero field resistivity ($\rho$) measurements in $LaFeAsO_{0.95}F_{0.05}$ is shown in Fig.1(a). The $\rho(T)$ curve shows a broad minimum at about 70 K and increases gradually with decreasing temperature down to $T_c$ ~ 26.3 K ($T_c$ is determined at which $\rho(T)$ equal to the 80% of the normal state resistivity). Such upturn was observed below ~10 % F-doping in LaFeAsO, which is substantially different from the metallic behavior in the heavily F-doped samples [1]. Figure 1(b) shows resistivity vs magnetic field $\rho(H)$. As illustrated in this figure, 60 T is insufficient to completely suppress the superconducting state below $T$ = 4 K. The field induced low temperature normal state resistivity shows positive magneto-resistance that increase with increasing magnetic field. This is markedly different compared to the recent high-field resistance measurement in $LaFeAsO_{0.89}F_{0.11}$ [6], in which the resistance at the highest field tends to saturate to a finite value. The change in temperature dependence with doping is reminiscent to the insulator-to-metal crossover observed in the normal state of high-$T_c$ cuprates. Indeed, underdoped and overdoped cuprates show insulating and metallic behavior, respectively [13, 14]. Shown in Figure 1(c) is the temperature dependence of resistivity at constant magnetic fields extracted from data in Fig. 1(b). The data clearly reveals the insulating behavior down to ~10 K, suggesting similar mechanism for insulating behavior in the underdoped cuprates and oxyarsenides. Although the upturn in the underdoped high-$T_c$ cuprates follows a log(1/$T$) behavior [14], the curve in our sample is better described with a $T^{-1}$ behavior above $T_c$ as shown in Fig.1(c) as a dashed line. Note that the temperature dependence below 20 K cannot be determined definitively from our data due to onset of the strong superconducting fluctuations.

Figure 2 plots the temperature dependence of the Hall charge carrier density ($n_H$) calculated from Hall coefficient ($R_H$); $n_H = (e|R_H|)^{-1}$. The Hall carrier density $n_H$ saturates close to ~5% per Fe atom at lowest temperature, which is consistent with nominal doping, and increases gradually with increasing temperature. The increase of $n_H$ with temperature is in sharp contrast to the temperature independent behavior expected in a normal metal, overdoped cuprates [13] and 11% doped LaFeAsO [10], indicating the striking thermal activation behavior of carrier in underdoped LaFeAsO. To further investigate the thermal activation behavior, assuming that the mobility remains the same for all carriers, we fit the data to the equation $n_H = n_0 + n_1\exp[-\Delta E/k_BT]$, where $n_0$ should correspond to the doping $x$, $n_1$ is the coefficient of the thermal activation term, and $\Delta E$ is the energy gap charactering the thermal activation behavior. These parameters are estimated to be $n_0 = 0.041 \pm 0.002$, $n_1 = 1.3 \pm 0.19$, and $\Delta E = 630 \pm 40$ K ($54 \pm 3.4$ meV ).

In some high-$T_c$ cuprates, a similar increase of $n(T)$ was interpreted as the thermal activation of holes due to a van Hove singularity in the density of states [15]. Thus, the activation behavior may indicate that the density of states has a peak structure within 54 meV from Fermi energy. The published band calculations in non-doped LaFeAsO show the peak in the density of states, but with a gap on the order of 250 meV [16]. While the simple comparison with non-doped LaFeAsO cannot account for the thermal activation behavior, the discrepancy may arise from the fluorine doping.

A plot of Hall angle, $\cot(\Theta_H) = \rho/R_H$, at 1 T versus $T^2$ is provided in Fig. 3. The increase of the Hall angle in the low temperature region is related to the upturn of the resistivity, where the temperature dependence of $R_H$ is small. Since $\cot(\Theta_H)$ corresponds to the inverse Hall mobility $\mu^{-1}$ at 1T, the increase in $\cot(\Theta_H)$ indicates a decrease of $\mu$. This suggests that the resistivity upturn below 70 K is due to decrease in $\mu$, most likely caused by disorder induced localization effects or spin fluctuation effects near the SDW instability. We note that the temperature dependence of $\cot(\Theta_H)$ seems to follow a $T^2$ dependence from 70 to 190 K. Although the $T^2$ temperature dependence is commonly observed in the underdoped high-$T_c$ cuprates [17], the $T^2$ law breaks down above the pseudogap temperature $T^*$ [18]. The $\cot(\Theta_H)$ in present data separate from $T^2$ dependence at about 190 K, which is well agreement with the $T^*$ of $210 \pm 14$ K calculated by the equation in Ref. 15, $T^* \sim -\Delta E / \ln x$. Therefore, we conclude that the activation behavior in $n_H$ is possibly originated from the opening of a pseudogap in the electronic density of states.

Our measurements of magnetoresistance $\rho(H)$ and TDO measurements, presented in Fig. 1(b) and the inset of Fig. 4, allow us to determine the $H_{c2}$ phase diagram. The open and closed circles in Fig. 4 are two characteristic fields which are at the onset of the $\rho(H)$ curves ($H_{onset}$) and the 80% value ($H_{c2}^{80}$) of the $T^{-1}$ fitting in Fig. 1(c). Since this system shows high anisotropy of $H_{c2}$ [19], the broad transition of $\rho(H)$ in Fig. 1(b) is due to the random grain orientation in polycrystalline sample. Thus, the $H_{c2}^{80}$ should correspond to the higher $H_{c2}$ in the $ab$ plane arranged parallel to magnetic field. On the other hand, $H_{onset}$ corresponds to the irreversibility field ($H_{irr}$) associated with the vortex liquid to solid transition [20]. In the analysis of TDO measurement, we determine the upper critical field ($H_{c2}^{TDO}$) deduced from bulk penetration measurements. The obtained $H_{c2}^{TDO}$ corresponds well to the average value of the $H_{c2}^{80}$ and $H_{onset}$ extracted from resistivity.

Figure 4 summarizes the characteristic magnetic fields in $H$ vs $T$ plane. For most superconductors,

WHH theory can quantitatively explain the temperature dependence of $H_{c2}$ [21]. The formula predicts a linear relationship near $T_c$, and is given by $H_{c2}(0) = -0.693(dH_{c2}/dT)_{Tc}T_c$. The observed $H_{c2}^{80}(T)$ is in good agreement with the WHH theory, yielding $H_{c2}(0) = 75.3$ T with $(dH_{c2}/dT)_{Tc} = -4.13$ T/K. This value is larger than the Bardeen-Cooper-Schrieffer (BCS) Pauli limit ($H_P$) of 48.4 T ($H_P = 1.84\ T_c$), as well as $H_{c2}$ in some high-$T_c$ cuprates which have higher $T_c$ ($La_{1.83}Sr_{0.17}CuO_4$; $T_c \sim 38$ K and $H_{c2} \sim 65$ T [20], $Bi_2Sr_2CaCu_2O_y$ $T_c \sim 80$ K and $H_{c2} \sim 65$ T [20]), and the recently estimated $H_{c2}$ in $LaFeAsO_{0.89}F_{0.11}$ ($T_c = 25 \sim 28$ K and $H_{c2} = 63 \sim 65$ T) [6]. The coherence length, $\xi_{GL}(0)$, obtained via Ginzburg-Landau (GL) formula ($H_{c2}(0) = \phi/(2\pi\xi_{GL}(0)^2)$ where $\phi$ is flux quantum, yields quite short coherence length $\xi_{GL}(0) \sim 21$ Å. These values suggest unconventional pairing mechanism in LaFeAsO system. Because the higher $H_{c2}$ corresponds to the stronger pairing potential [22], this indicates that the excess fluorine doping weakens the paring potential. Such doping dependence of $H_{c2}$ has been reported in some high-$T_c$ cuprates [23].

## Summary

The present study of resistivity and Hall coefficient in the new superconductor $LaFeAsO_{0.95}F_{0.05}$, uncovered some similarities with the high-$T_c$ cuprates in the underdoped region. These are an insulating-like behavior in $\rho$, a $T^2$ dependence in $\cot(\Theta_H)$, and a thermal activation behavior in $n_H(T)$. The magnitude of the estimated activation energy $\Delta E$ may relate to the pseudogap, and the next step will be to elucidate a correlation between the $\Delta E$ estimated from $R_H$ and $T^*$ determined by the other measurement techniques. Additionally, the quite large value of $H_{c2}$ (i.e., short coherence length) suggests an unconventional pairing mechanism. Further studies over a wide doping range are presently under way.

## Acknowledgements

This work was supported by Grant-in-Aid for JSPS Fellows (No. 19·9728), the DOE, the NSF, and Florida State University through the National High Magnetic Field Laboratory. We would like to thank H. Kawaji, H. Yuan and J. Singleton for technical assistance during the experiments and helpful discussion.

Figure captions

Figure 1 Resistivity measurements: (a) Temperature dependence of resistivity, $\rho(T)$, at zero field. (b) Magnetic field dependence of resistivity $\rho(H)$. (c) The temperature dependence of $\rho$ under various magnetic fields. The dashed curve is the fit of $T^{-1}$ (see the text).

Figure 2 Temperature dependence of Hall charge carrier density ($n_H$). The dashed curve is the best fit of the fitting equation (see the text).

Figure 3 Temperature dependence of Hall angle ($\cot(\Theta_H)$). It is determined from the $\rho(T)$ data in Fig. 1(a) and $R_H$. The dashed line shows the best fit to the $T^2$ dependence of $\cot(\Theta_H)$ in the 70 – 170 K, indicated by arrows.

Figure 4 Main Panel; Temperature dependence of characteristic fields ($H_{c2}^{80}$, $H_{onset}$, and $H_{c2}^{TDO}$). The open circles, close circles, and squares are $H_{c2}^{80}$, $H_{onset}$, and $H_{c2}^{TDO}$, respectively. These values are evaluated from the data in Fig. 1(b) and in the TDO measurement. The dashed curve corresponds to the theoretical WHH curve (Ref. 21). Inset; Result of the TDO measurement. The $H_{c2}^{TDO}$ values are estimated by the point at the intersection of the high-field extrapolation with low-field extrapolation.

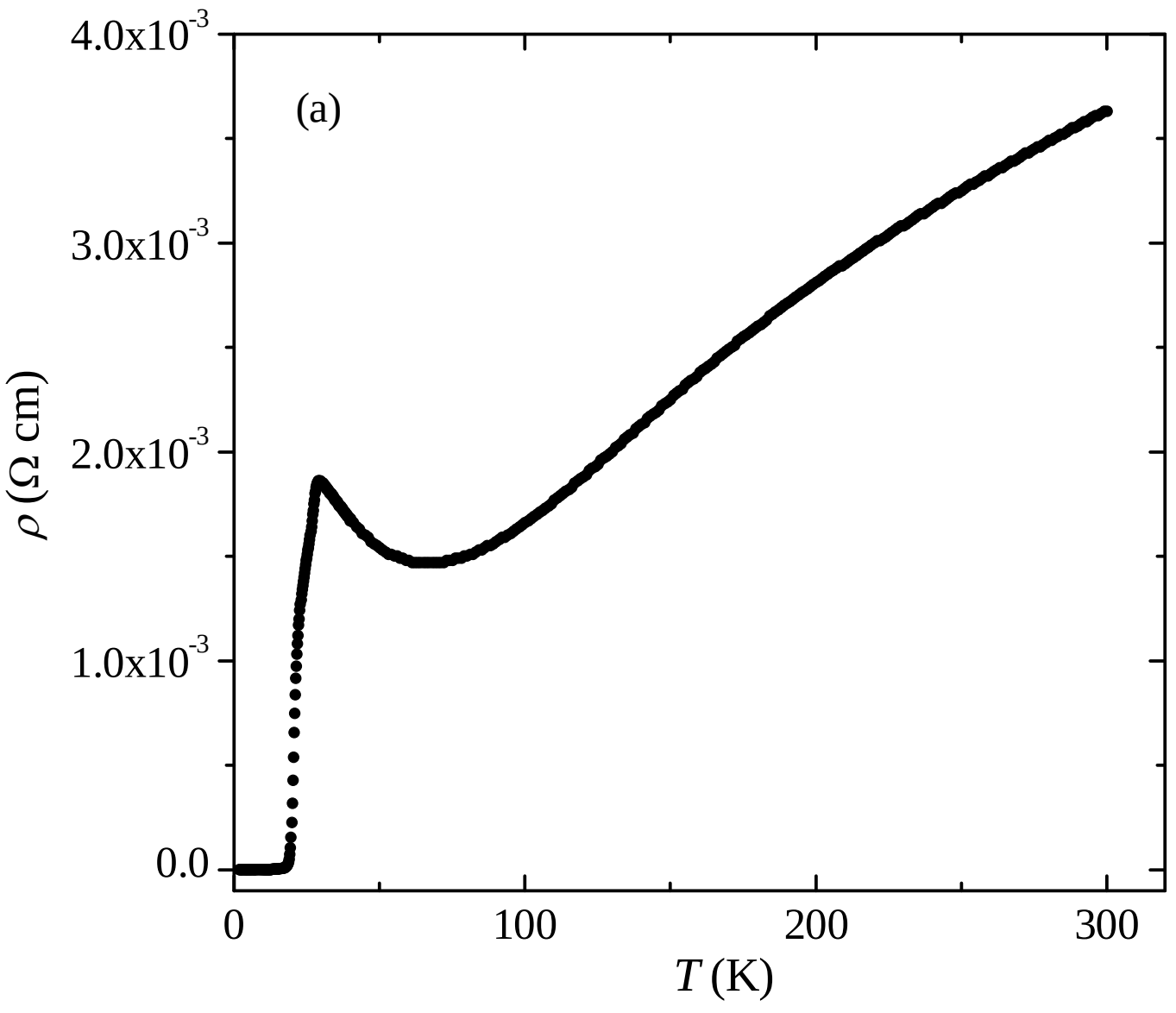

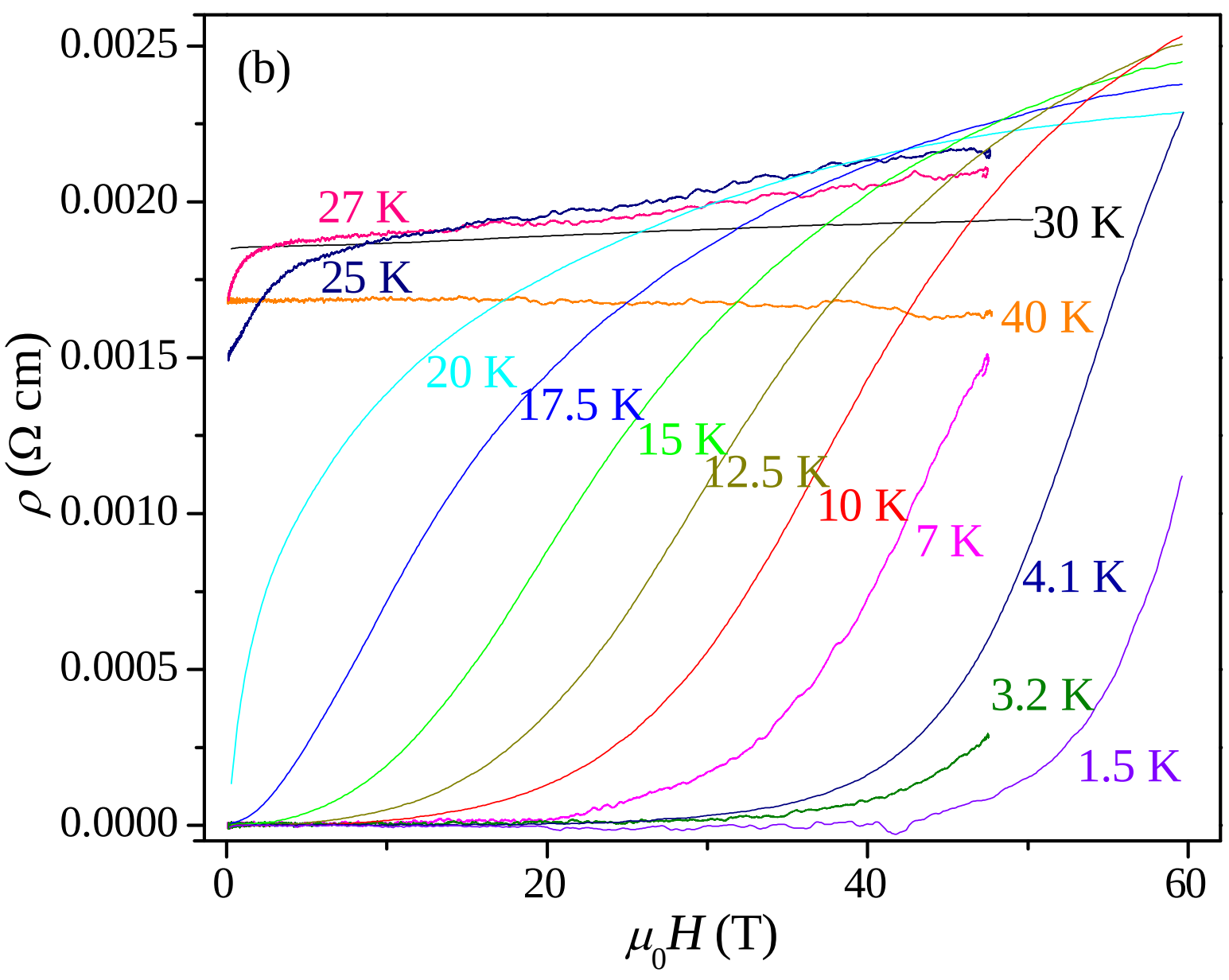

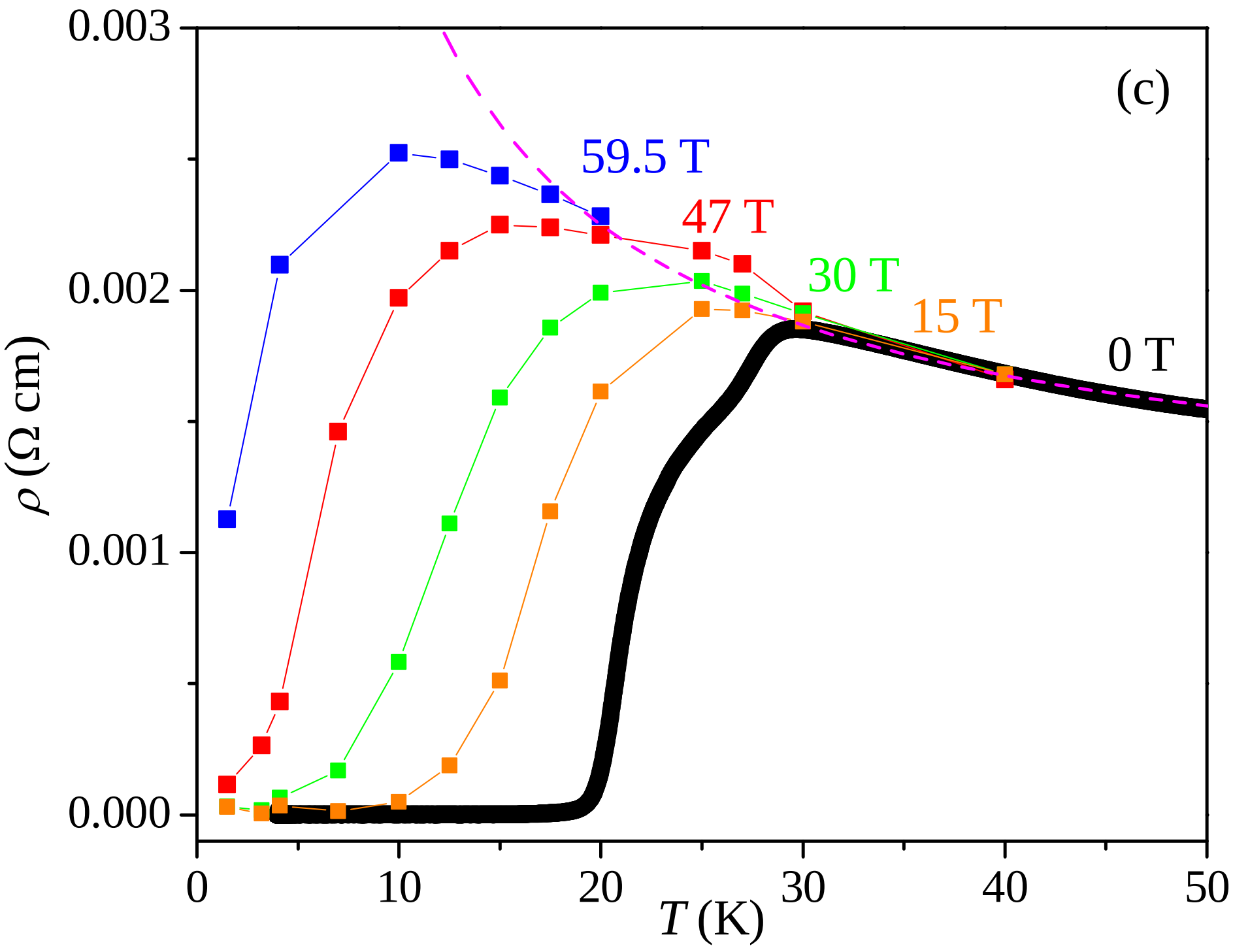

Fig. 1. Y. KOHAMA, et al.

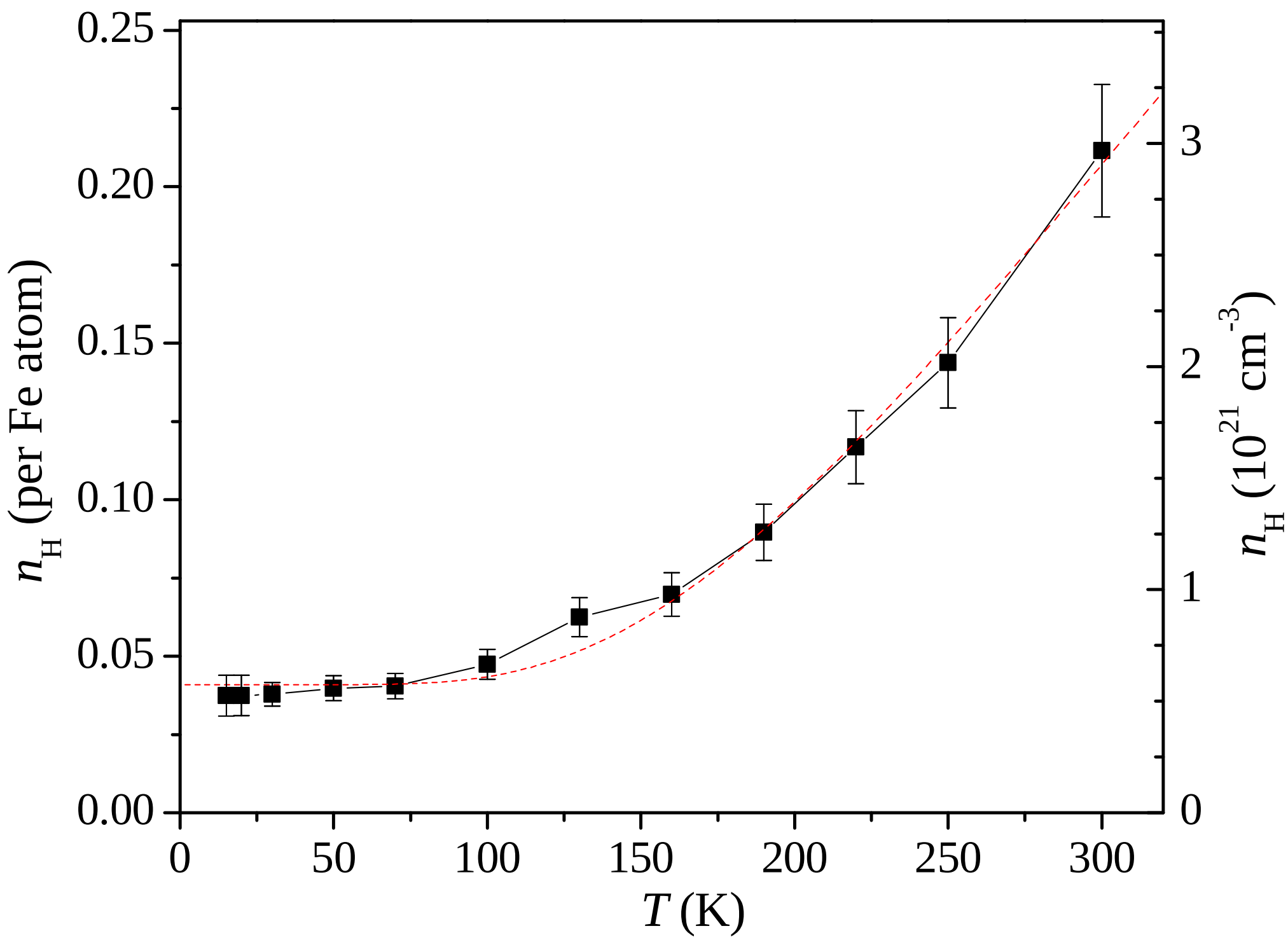

Fig. 2. Y. KOHAMA, et al.

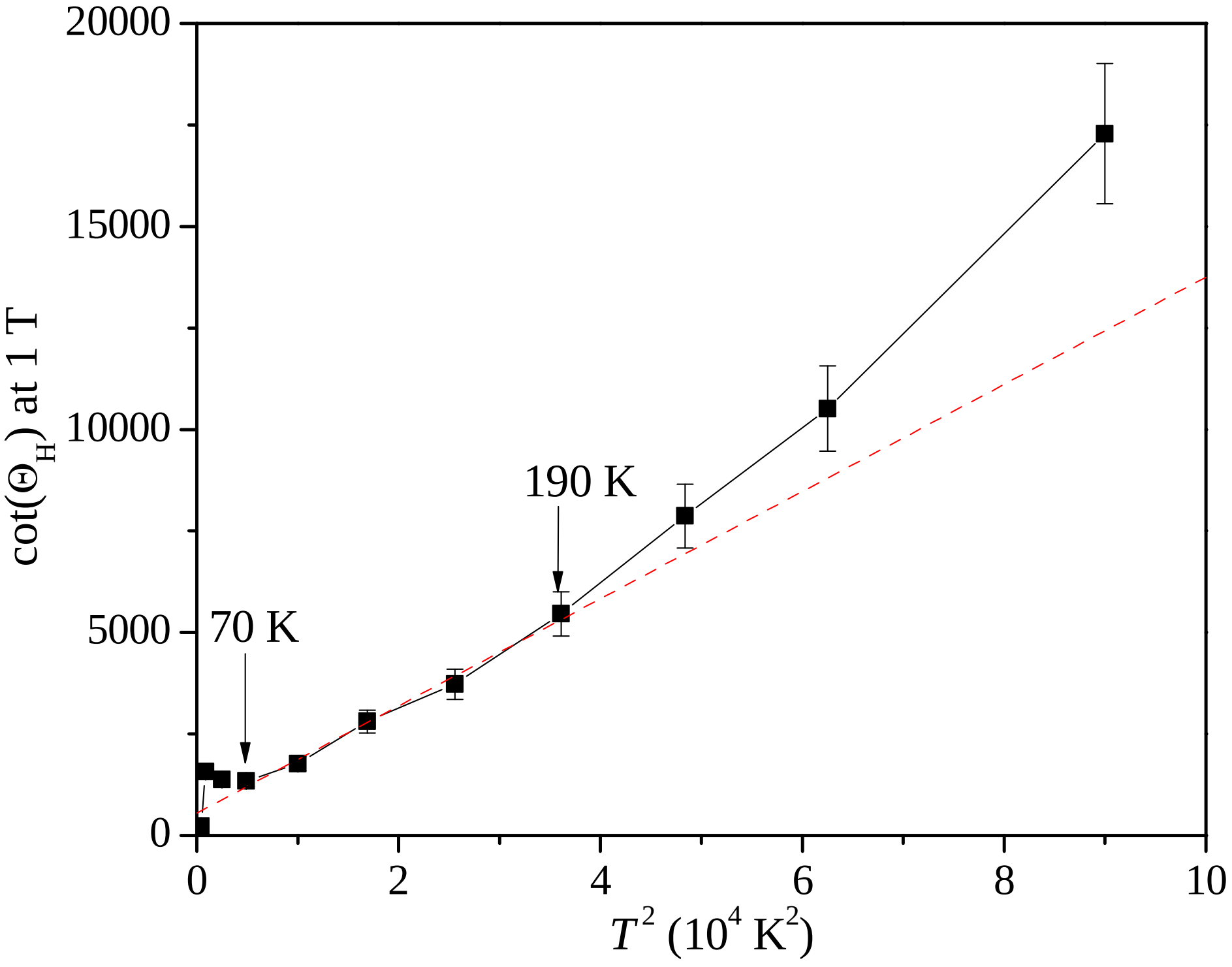

Fig. 3. Y. KOHAMA, et al.

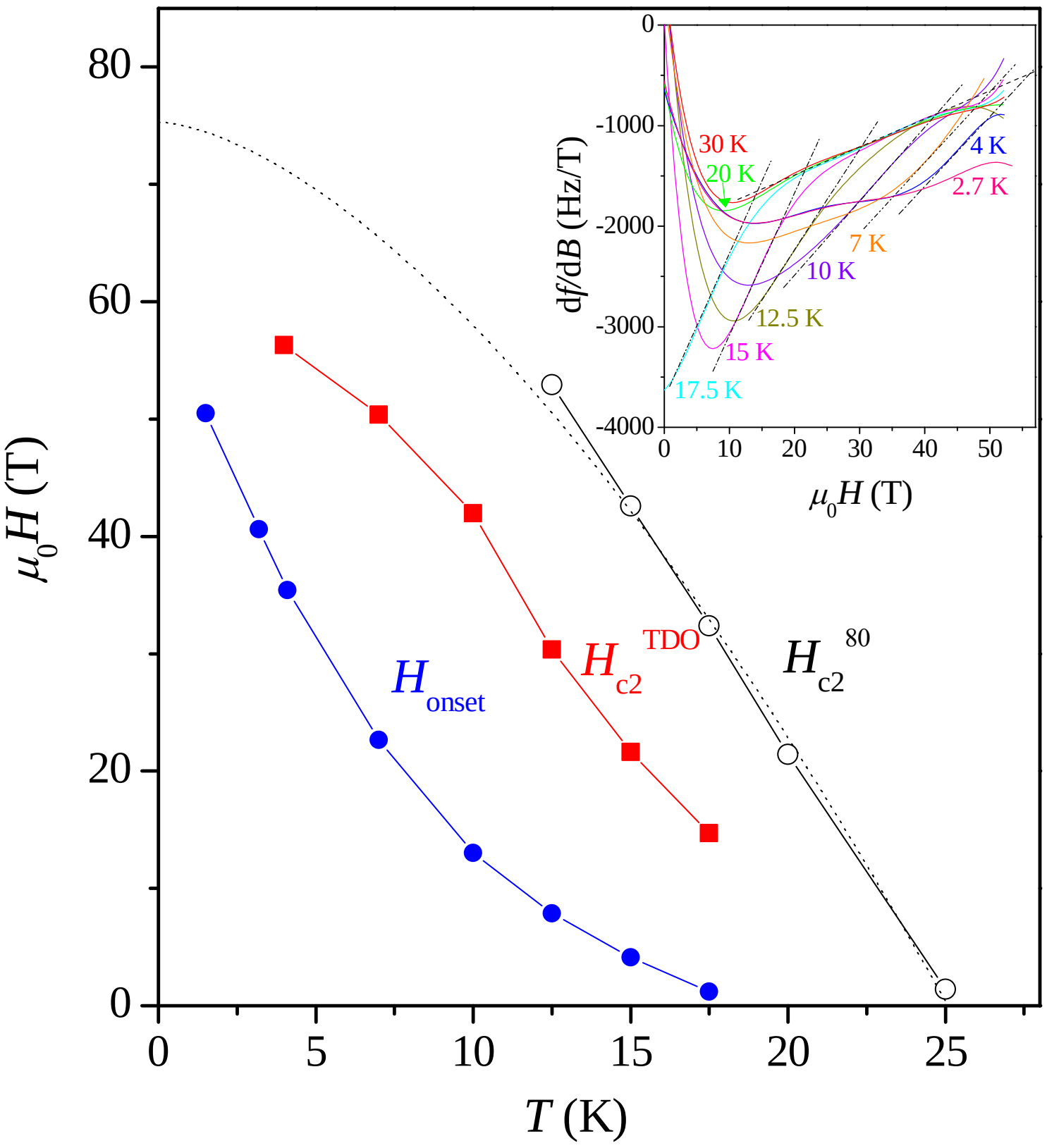

Fig. 4.     Y. KOHAMA, et al.